****
# FRIEDMANN THERMODYNAMICS AND THE GEOMETRY OF THE UNIVERSE

**Selçuk Ş. BAYIN**
Department of Physics
Middle East Technical University
Ankara-06351
TURKEY

**ABSTRACT**

In a recent article we have introduced Friedmann thermodynamics, where certain geometric parameters in Friedmann models are treated like their thermodynamic counterparts (temperature, entropy, Gibbs potential etc.). This model has the advantage of allowing us to determine the geometry of the universe by thermodynamic stability arguments. In this article we review connections between thermodynamics, geometry and cosmology.

**GEOMETRY AND THERMODYNAMICS**

After the initial success of the Einstein's general theory of relativity in explaining the perihelion shifts of Mercury and the deflection of light caused by the spacetime curvature of the Sun, it was immediately applied to cosmology and gave birth to the Friedmann models. These are homogeneous and isotropic models and have successfully incorporated the expansion of the universe, explained why the night sky is dark and predicted the existence of the universal background radiation.

One of the distinct features of the Friedmann models is that they give three possible geometries for the universe, which could be represented by the following line element [1];

$$ds^2 = dt^2 - R^2(t)\left[dr^2/(1-\mathbf{k}r^2) + r^2 d\theta^2 + r^2 \sin^2\theta\, d\phi^2\right]. \quad (1)$$

If we consider constant time slices (t=const.) of the spacetime geometry represented by this metric, we see that for various values of the constant k it gives three possible geometries for the space on which the galaxies are distributed. Zero value for **k** corresponds to flat space of infinite extent, where the underlying geometry is Euclidean. For **k**=1 models, the physical space in which we live corresponds to the (three dimensional) surface of a 4 dimensional hypersphere, where the underlying geometry is Gaussian. Models with **k**=-1, correspond to spaces with negative curvature and their geometry is called the Lobachevski geometry. R(t) in our line element (1) is the scale factor and is a measure of how fast the universe is expanding. For **k**=+1 models, R(t) also corresponds to the radius of the hypersphere. Despite the simplicity of the Friedmann models, it is commonly accepted that they present a fairly accurate description of the global properties of the universe.

Given an equation of state for the matter content of the universe, Einstein's field equations could be used to determine the expansion factor R(t). However, the value of **k** (+1, -1, or 0) which determines the type of the geometry is left as a boundary condition that could only be determined by observations. Today, even though this problem is still far from being settled, there exists a strong evidence for the **k**=-1 case (Gott, 1982, Boesgaard et.al., 1985).

In thermodynamics it is well known that different spatial distributions (crystal structures) of atoms have different entropies. Hence, at a given temperature the second law of thermodynamics allows one to predict the crystal structure that is favored by nature. Depending on the discontinuities in the derivatives of the Gibbs potential, changes in the crystal structure at the critical temperature are identified as first -or second-order phase transitions (Landau et. al., 1974). A typical example is furnished by the α- and γ-iron case. α-iron is also known as ferrite and has body centered cubic (bcc) crystal structure, while γ-iron is austenite and has face centered cubic (fcc) crystal structure. As we increase the temperature of α-iron, at a certain critical temperature it becomes thermodynamically advantageous for the iron atoms to redistribute themselves over an fcc type lattice and the crystal structure begins to change. This transition between two solid phases of iron is a first-order phase transition just like melting, freezing or boiling.

---

[1] Line element is what replaces Pythogoras' theorem in curved spacetimes. It gives the square of the distance between two infinitesimally close points. Presence of curvature manifests itself as some of the coefficients of the coordinate differentials (metric tensor components) as being functions of coordinates (note that the reverse statement is not true).

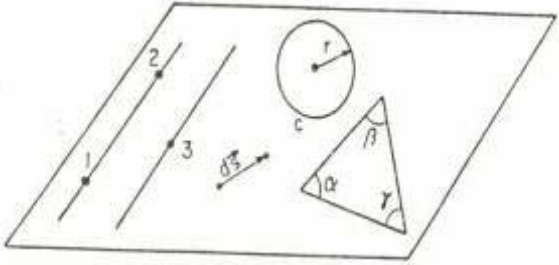

**GEOMETRY OF SPACE** *There is a surface characteristic of each model universe. The properties of the surface are defined by the Euclidean axioms and theorems on parallel lines, on the interior angles of triangles and on the circumference c and area A of a circle. For **k**=0 flat universes $c = 2\pi r$, $A = \pi r^2$, $\alpha+\beta+\gamma = \pi$ and there exist only one parallel line to a given line through given point 3. Shortest distance between two points (geodesics) is given by straight lines.*

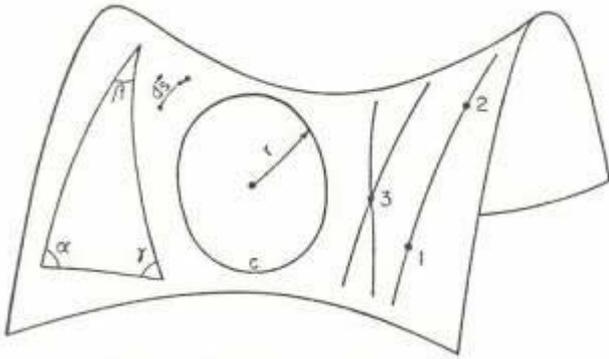

**NEGATIVELY CURVED SURFACE** *Some of the properties of the negatively curved space (**k**=-1 models) could be represented on a saddle shaped surface. In Lobachevski geometry $c > 2\pi r$, $A > \pi r^2$, $\alpha+\beta+\gamma < \pi$ and there exist more than one parallel to a given geodesic. Keep in mind that saddle is an imperfect analogue since it has a center.*

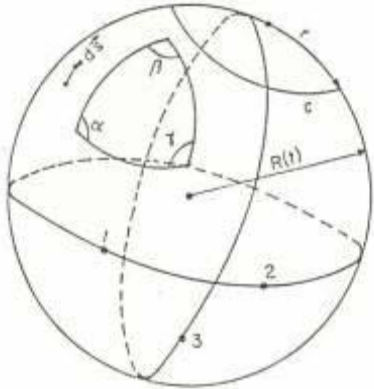

**POSITIVELY CURVED SURFACE** *The shortest path between two points follows great circles, while $c < \pi r$, $A < \pi r^2$, $\alpha+\beta+\gamma > \pi$. There are no parallel lines, since any two great circles necessarily intersect.*

The fact that in Friedmann models, Einstein's theory can not determine the geometry of the universe but gives three alternatives (**k**=0 or ±1) is deeply rooted in the fact that all three models are isentropic and have the same microscopic equation of state $P = P(\rho)$. Hence, the type of the geometry is left as a boundary condition to be determined by observations. Notice that the distribution of galaxies over different spatial geometries is very similar to the distribution of atoms over different lattices. In lattices there exists certain discrete crystal symmetries while the galaxies in Friedmann models are distributed over spaces with distinct but this time continuous symmetry properties classified as Bianchi symmetries. One could note that in the case of α- or γ- iron we have a classical system where the clocks carried by all the atoms run at the same rate, while the Friedmann models are relativistic models with the curvature in question being the curvature in four dimensional spacetime. However, in Friedmann models the time coordinate is still the classical time, i.e. clocks carried by all the galaxies run at the same rate. This is evident in the line element by the fact that the coefficient of $dt^2$ is 1, and that there are no cross terms like $dtdr$, $dtd\phi$, and $dtd\theta$. For this reason the geometries that we have been discussing actually correspond to the geometry of the three dimensional physical space that we live in.

Motivated by these similarities we have recently considered Friedmann models from a different point of view and discussed the possibility of predicting the geometry of the universe by entropy arguments (Bayin, 1986, 1987, 1990, 1994). Within the standard view of the Friedmann models we have seen that the type of our geometry is left as a boundary condition to be determined from observations. Thus, we need to find some new kind of entropy and temperature such that when this new entropy is considered along with the existing entropy, it would give us a way to compare different geometries thermo-dynamically.

Our starting point was first to look for certain geometric parameters in Friedmann models that behaves like their thermodynamic counterparts i.e. temperature, entropy, Gibbs energy etc. We first looked for a geometric quantity that has the properties of temperature i.e. a quantity that is uniform through out the system and a scalar with respect to spatial coordinate transformations. The absolute value of the inverse of the radius of the hypersphere $|1/R(t)|$ has the desired properties. Notice that flat (**k**=0) models could be considered as hyperspheres with infinite radius, while negatively curved (**k**=-1) models could be viewed as hyperspheres with imaginary radius. Identifying this quantity as proportional to the "gravitational" temperature we were able to extract an expression that was interpreted as the Gibbs energy density of the system. Hence, for a given "local" equation of state we were able to compare the three geometries in terms of the thermodynamics that we have defined. In fact for a local equation of state given as $P=\alpha\rho$ ($\alpha$ is a constant) we were able to deduce that the value of **k** should be -1, which is in line with the current observations. One advantage of our method is that it allows the universe to undergo symmetry (geometry) changes as it evolves and such symmetry changes are interpreted as first-or second-order phase transitions.

Despite the interesting and promising nature of the Friedmann thermodynamics, we have to realise that behaviour of certain geometric parameters like some thermodynamic variables could be just a coincidence. Thus, for our arguments to have physical relevance we have to find some evidence to the effect that these quantities are also physically related to their thermodynamic counterparts.

In this regard we have recently studied the quantum vacuum energy of the massless conformal scalar field in static closed

Friedmann universe (also called Einstein universe) Bayin (1990,1994). We argued that the " gravitational " temperature of this universe is given as T=(1/2π) ($\hbar$ c/k) (1/$R_0$), where $R_0$ is now the constant radius of the universe: Thus determining the only arbitrary constant in Friedmann thermodynamics as (1/2π). For Friedmann models, granted that they are expanding sufficiently slowly the "gravitational" temperature will be given by[2]

$$T=(1/2\pi)(\hbar c/k)(1/R(t)). \qquad (2)$$

We would like to point out that in standard Friedmann models we compare three models with different geometry, different expansion factor but the same equation of state P = P(ρ). The different expansion factors are found by solving the Einstein's field equations for a given equation of state and for the three possible values that **k** could take. Thus, these models are isentropic and can not be distinguished entropy wise. On the other hand, in Friedmann Thermodynamics we compare three different geometries with the same expansion factor. These three geometries have the same local equation of state but differ in terms of their "gravitational" entropy. The local equation of state could be determined by studying a region large enough to contain many galaxies but not large enough for the effects of curvature to be important. In practice it corresponds to the equation of state P = P(ρ) that microscopic physics gives.

**Is Absolute Zero Really Zero ?**

In 1912 Nernst pointed out that as a consequence of the third law of thermodynamics, it is impossible to cool any system down to absolute zero. He argued that by successive steps of isothermal compression followed by an adiabatic expansion, one should be able to approach absolute zero as close as one desires, thus explaining the unattainability of absolute zero by the fact that one would need infinite number of such steps (Wilks, 1961). Among physicists this immediately set up a tantalizing quest for absolute zero. Considering that the temperatures between $10^{-6}$K - $10^{-9}$K etc. could span a variety of physical phenomena comparable to the 1 K - $10^3$ K range, the race for the lowest temperature became a popular avenue of research that will maintain its important status for many years to come. The latest developments in this race has been discussed by Lounasmaa (1989) where he has reported a new record low temperature at $2\times10^{-9}$ K.

Another interesting problem that has tantalized physicists and philosophers is the creation of absolute vacuum in a given region of space. In a recent popular article Boyer (1985) has discussed the classical vacuum concept starting from Aristotle. An obvious way to create a vacuum is to remove all visible matter such as solids, liquids, as well as gases. With the revolutionary work of Planck and the development of quantum mechanics in the early twentieth century, it became clear that such a region will still be not empty. Infact, it will contain thermal radiation at the temperature of the walls. Hence, in order to create vacuum in a given region, evacuating that region from all forms of matter should also be accompanied by refrigeration of the walls to absolute zero.

With the development of quantum field theory and the beautiful work of Casimir (Plunien et.al., 1986) it became clear that even if we could eliminate the thermal radiation (at least in principle) by cooling the walls to absolute zero, the region will still contain some radiation called zeropoint radiation. Casimir's initial calculation was for the region between two infinitely large conducting plates and he found out that due to the zeropoint energy there exist a small attractive force between the plates. The fact that Casimir effect is real and has to be taken seriously became evident by the experiments of Sparnaay (Plunien et.al., 1986). Casimir's calculations were later extended to other cases, where the boundary is defined by a cube and a sphere (Plunien et.al., 1986). Contrary to the parallel plate case the force exerted on these new boundaries by the zero point radiation turned out to be repulsive. This shattered some early hopes of constructing an electron model where the Casimir effect would be used to counterbalance the repulsive electrostatic forces. One important feature of the zeropoint radiation is that it corresponds to a pure state. Thus, the entropy and the temperature associated with it are zero. In realistic situations there are always interactions no matter how small they may be. Hence, a pure state is eventually expected to thermalize. In the original work of Casimir we have a rather idealistic model, where the walls are at absolute zero and the field under consideration is noninteracting, hence the zeropoint radiation always remains in a pure state.

Another blow to our classical concept of vacuum came with the work of Unruh (1976). His results could be summarised by saying that what an inertial observer calls a vacuum state (a state defined by the absence of real particles) is perceived by a uniformly accelerating observer as a thermal bath of particles at the temperature T=($\hbar$/2πkc)a, where a is the acceleration of the observer. One more intriguing result follows from Unruh's work. Using the equivalence principle which relates an observer at rest in a gravitational field with a uniformly accelerating observer, Boyer (1985) concluded that there may be a minimum attainable temperature on Earth. Replacing the acceleration in Unruh's formula with the gravitational acceleration at the surface of the Earth ($GM_\oplus / R_\oplus^2$) he obtained $4\times10^{-18}$ K. Unfortunately this is far below the current record of lowest temperatures reached in the laboratory. Besides , even though such simple uses of the equivalence principle may help to make a point, it may also lead to a wrong result. In fact, calculation of the deflection of light by the Sun through the use of the Newtonian potential and the principle of equivalence gives us a result which is off by a factor of two. Since the field of a star is not uniform, one has to make use of the exact solution of the appropriate relativistic field equations.

Using the definition of gravitational temperature that we have obtained in Friedmann thermodynamics (2) we could obtain a temperature that may have a better chance of being correct. Notice that even though the temperature given in (2) is derived as a zero temperature and zero entropy state, it differs from the ordinary Casimir case by the presence of interactions. The scalar particles interact with the background geometry. Thus, we expect them to thermalize and the temperature given in (2) represents the temperature at which they will thermalize. For this reason we have identified it as the "gravitational" temperature of the

---

[2] $\hbar$= Planck constant, c=speed of light, k=Boltzman constant. Since $\hbar$ appears in the definition of "gravitational" temperature, it is intrinsically a quantum effect.

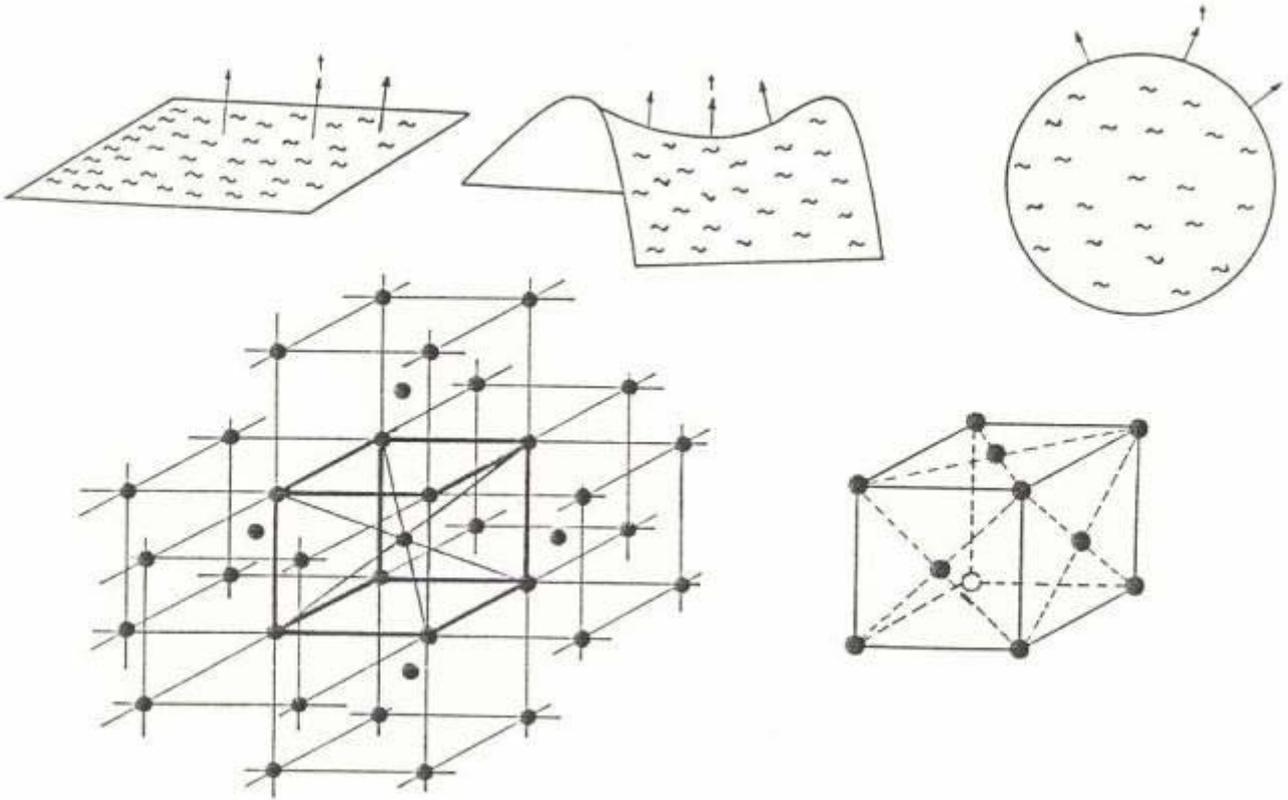

**GALAXIES VERSUS ATOMS** *In Friedmann models galaxies are distributed over continuous spaces with distinct symmetry properties given by Bianchi symmetries. In crystals atoms are distributed over lattices with certain discrete crystal symmetries. In Friedmann models the fourth dimension time is classical and orthogonal to the space on which galaxies are distributed.*

universe. At the quantum level the thermalization process could be understood as due to the random scattering of scalar particles by the background gravitons. The possibility of a thermal interpretation of the Casimir energy in an Einstein universe, even though in a slightly different context, has also been noticed by others (Mostapanenko et.al., 1988, Bunch et.al., 1977). Despite the plausibility of the thermalization argument a rigorous derivation of the "gravitational" temperature and details of how thermalization may take place is not known. Besides, the approximate methods of quantum field theory on curved background spacetime which we have used to derive (2) may not be sufficient to answer these questions. In the above picture the background geometry acts like a heat bath at the temperature given in (2). A vacuum state which has zero temperature on flat spacetime acquires a temperature when considered on curved background spacetime. Naturally, this temperature (2) defines a new lowest temperature in an Einstein universe. Since gravity is a universal interaction, it should also be a universal limiting temperature regardless of the field used. When we consider systems at finite temperature on a curved background geometry, the "gravitational" temperature and the original temperature will remain as properties of two weekly interacting components of the system. How they affect each other is beyond the scope of the background field approach (Bayin, 1990).

It is well known that in thermodynamics temperature is well defined only for systems in thermal equilibrium. For this reason Friedmann models are very suitable for our purpose and allow us to define a homogeneous temperature throughout the system. In generalizing our definition of temperature to more general spacetimes we recall that, in thermodynamics, systems that can not be represented by rigorous thermodynamic equilibrium may still permit the definition of a temperature which describes the local properties of the system to a high degree of accuracy. In such cases we basically say that the material in the neighborhood of some point P is in "local thermodynamic equilibrium" and define a temperature that is a function of position and possibly time. Hence, in general relativity we may still be able to assign a local "gravitational" temperature at each point, granted that spacetime geometry changes sufficiently slowly both with space and time. In particular, we recently considered the Schwarzschild geometry (Bayin,1990) and by taking its constant time slices we approximated the resulting spatial geometry (in the neighbourhood of a given point P) by the surface of a hypersphere. We used the radius of this sphere to suggest a local gravitational temperature, which is now a function of radial position r as

$$T = (1/4\pi)(\hbar c/k)\left[(2GM)/(c^2 r^3)\right]^{1/2} \cdot \quad (3)$$

An interesting result emerges when we consider the gravitational temperature at the surface of a star and take the blackhole limit i.e. as the surface of the star approaches to its horizon ($R_{surface} \rightarrow 2GM/c^2$). In this limit the "gravitational" temperature at the horizon becomes

$$T_{horizon} = (1/8\pi)(\hbar c^3/Gk)(1/M), \quad (4)$$

which is precisely the Hawking temperature for blackholes. For experiments on Earth the meaningful limiting temperature would naturally be given by (3). Substituting the mass and the radius of

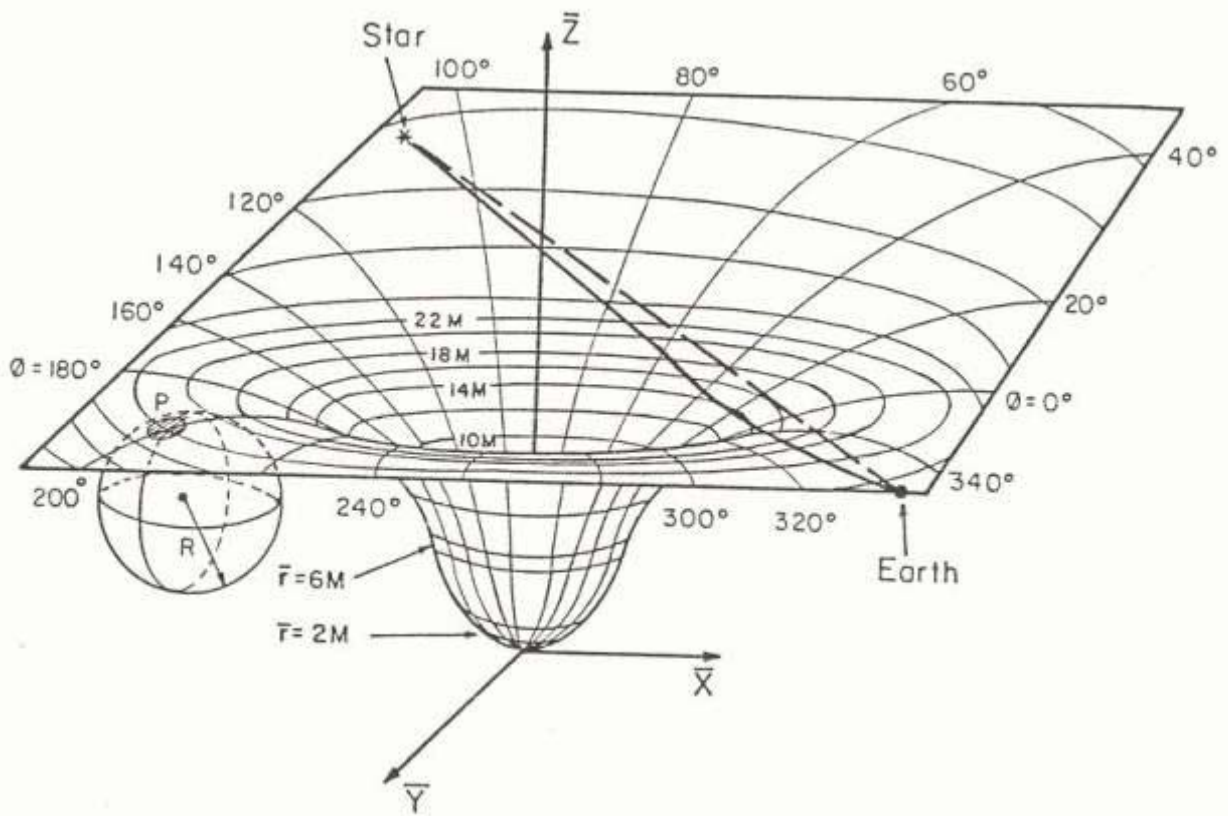

**SPACE IS CURVED IN THE PRESENCE OF MASS** *The curvature is depicted in what is known as an embedding diagram, in which three-dimensional space is represented by a flat plane that is warped by the presence of a star or any other kind of massive object. The amount of curvature is related to the objects mass. In this picture we present the embedding diagram of the Schwartzschild solution, which gives the geometry of space near a spherically symmetric star of mass M (e.g. Sun). Deflection of light is shown here. In the presence of the Sun, starlight follows the solid geodesic to Earth, while when the Sun has moved out of sight the light follows the dotted line. The sphere to the left represents the hypersphere of radius R, whose surface approximates the Schwarzschild geometry in the neighbourhood of the point P.*

the Earth we get $10^{-15}$ K. This is 3 orders of magnitude larger than the Boyer's result and has a better chance of being detected in future.

In thermodynamics it is well known that systems with finite temperature are also endowed with a finite entropy. Hence, it is natural to expect that the background geometry, which acts like a heat bath at the "gravitational" temperature T, also has a finite "gravitational" entropy. Since this "gravitational" entropy will define a new universal origin for the entropy of a given thermodynamic system considered on curved background spacetime, the lowest temperature given by (2) will also be unattainable but only asymptotically approachable. Existence of "gravitational" entropy could be understood by remembering the fact that gravity is a universal interaction and couples to the total energy-momentum tensor with a universal coupling constant. Hence, by looking at a gravitational field all we can say about its source is its total energy-momentum distribution. However, there are many ways to produce a given energy-momentum distribution. This naturally reminds us the definition of entropy in statistical mechanics, where it is defined as proportional to the logarithm of the number of microstates that belongs to the same macrostate. This may suggest an analogous definition of "gravitational" entropy as proportional to the logarithm of the number of different sources that will lead to the same energy-momentum tensor, hence to the same spacetime geometry. In classical gravity this number is infinity. However, it could be seen that quantum mechanics will help to get a large but nevertheless a finite number. Unfortunately, since we do not yet have the underlying statistical mechanics to Friedmann thermodynamics we can not calculate this entropy.

Behaviour of blackhole parameters (area, surface gravity and so on) like certain thermodynamic quantities has motivated Bekenstein (1980) to conjecture blackhole thermodynamics. Later, the discovery of blackhole radiation established the physical link between these parameters and their thermodynamic counterparts. Despite the success of blackhole thermodynamics, relation between geometry and thermodynamics has to be established for general spacetimes (Bonner, 1985). Friedmann thermodynamics may help us to bridge this gap. Notice that in establishing the physical link between geometry and thermodynamics the calculations that are used are based on the approximate methods of quantum field theory on curved background geometry i.e. gravity is still considered as a classical field. At the early stages of the development of the quantum electrodynamics we have experienced that some of the results that emerged from such approximate theories may surprisingly remain to be valid even in the exact theory. However, in this case the difficulty of our task could be understood by the fact that it is not even clear what is to be expected from the "exact" theory and what is to be understood from "quantum" geometry. So far all attempts that have calmly ignored the geometric side of the Einstein's field equations, hence treated gravity as just another field and tried to quantize it, have failed to produce a satisfactory picture. Surely, the difficulties are not just technical. Exploratory models like Friedmann thermodynamics are instrumental in giving us new lookout points to the universe, where we could get some ideas on the whereabouts of the missing concepts that may take us to the next generation of theories in science. Whatever this new theory is, we suspect that it is somewhere at the intersection of geometry, thermodynamics and quantum mechanics. It has the potential to change our views of the micro- and macro-cosmos, while bringing them together under a new and a broader definition of observer and its relativity.

## ACKNOWLEDGEMENT

I would like to thank Prof. J.Krisch for a critical reading of the manuscript.

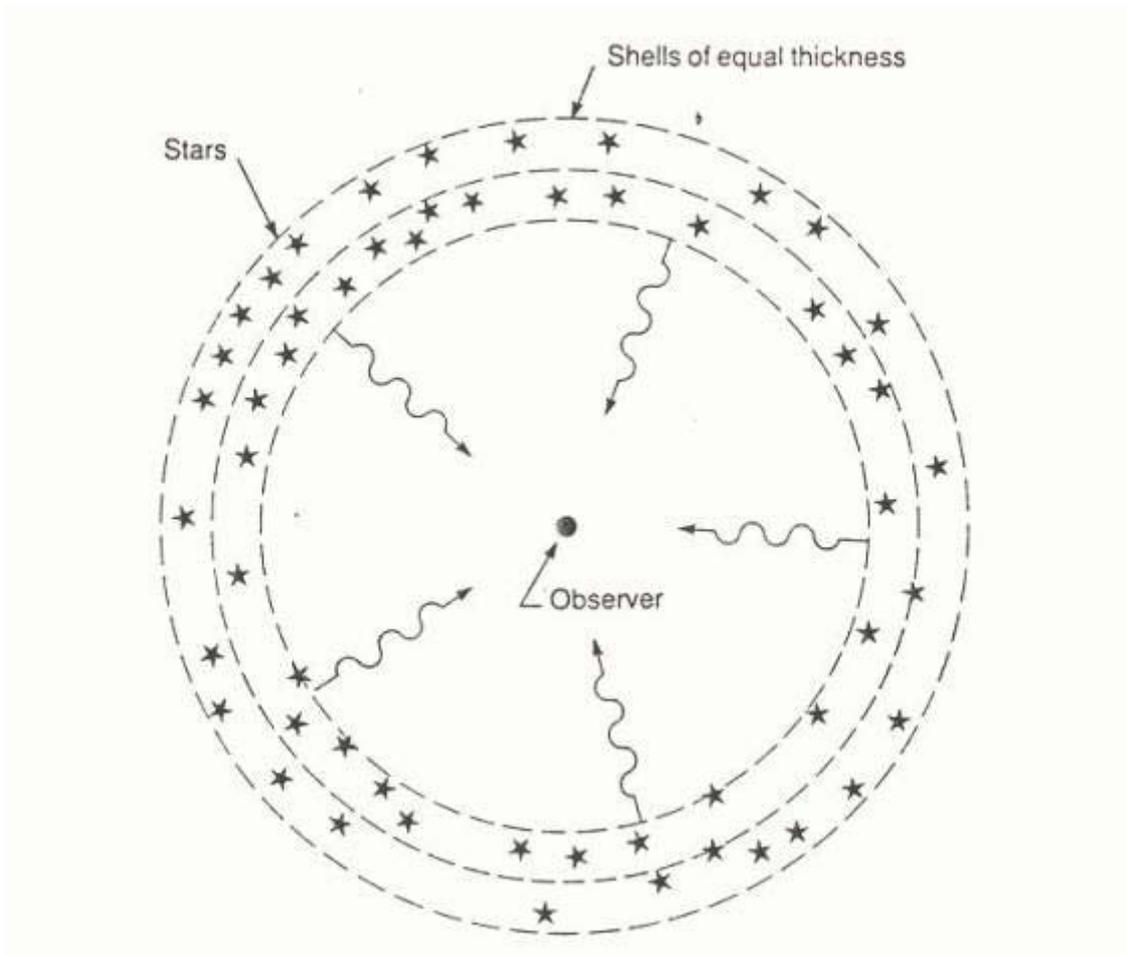

**WHY THE NIGHT SKY IS DARK ?** *If the universe is unchanging with time and uniformly populated with stars similar to the Sun to infinity, we encounter a remarkable paradox named after Olber. For whatever direction we look in the sky, our line of sight eventually intercepts a star. Thus, the night sky should be ablaze with light as bright as the Sun. Notice that as we consider shells of increasing radius brightness of the stars decrease as 1/r². However, since the number of stars in each shell increases as r² the flux remains constant.*

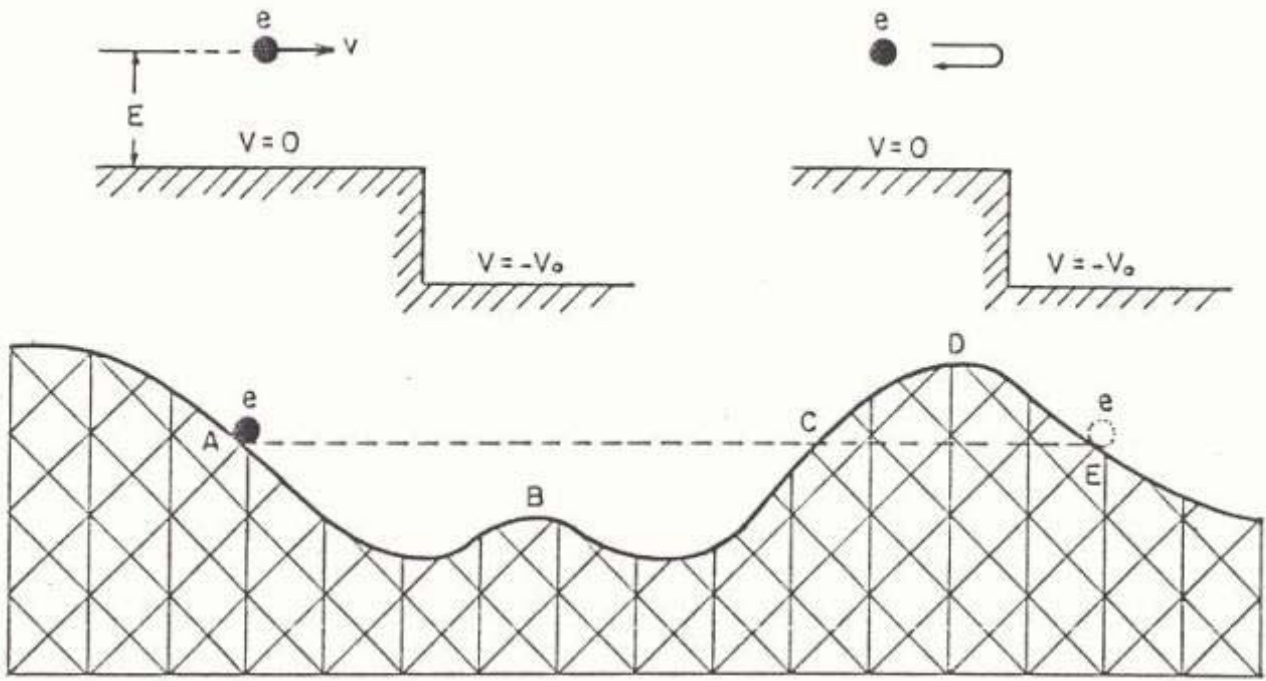

***IN THE MICRO-COSMOS STRANGE THINGS HAPPEN*** *An electron travelling towards a cliff (potential drop of $V_0$) has a finite probability of getting reflected at the edge. In a roller coaster conservation of energy forbids a particle that starts from point A with zero velocity from overshooting the point C. However, in quantum mechanics it is possible for an electron to sneak through the forbidden region between C and E, and emerge on the other side of the hill. Such events have many interesting practical applications like the scanning tunnelling microscope. The fact that $\hbar$ appears in the definition of "gravitational" temperature makes it intrinsically a quantum effect, thus impossible to understand interms of classical concepts.*